\documentclass{article}
\usepackage{emulateapjnew,apjfonts,rotating,epsf}                                        

\lefthead{Wuchterl \& Klessen}
\righthead{The first million years of a solar mass star}

\begin{document}

\title{The first million years of the Sun: A calculation of formation
  and early evolution of a solar-mass star}
\author{G.\ Wuchterl\altaffilmark{1,2} \& Ralf S.\
Klessen\altaffilmark{3,4,2}}

\altaffiltext{1}{Max Planck Institut f\"{u}r extraterrestrische Physik,
                            P.O Box 1312, D-85741 Garching
                            Germany}
\altaffiltext{2}{Sterrewacht Leiden, P.O. Box 9513, NL--2300 RA,
                            Leiden, The Netherlands}

\altaffiltext{3}{UCO/Lick Observatory,
                            University of California at Santa Cruz,
                            499 Kerr Hall, Santa Cruz, Ca 95064, USA}
\altaffiltext{4}{Max-Planck-Institut f{\"u}r
 Astronomie, K{\"o}nigstuhl 17, 69117 Heidelberg, Germany}

%


\begin{abstract}
   We present the first coherent dynamical study of the cloud
   fragmentation-phase, collapse and early stellar evolution of a
   solar mass star. We determine young star properties as the
   consequence of the parent cloud evolution.  Mass, luminosity and
   effective temperature in the first million years of the proto-Sun
   result from gravitational fragmentation of a molecular cloud region
   that produces a cluster of prestellar clumps. We calculate the
   global dynamical behavior of the cloud using isothermal 3D
   hydrodynamics and follow the evolution of individual protostars in
   detail using a 1D radiation-fluid-dynamic system of equations that
   comprises a correct standard solar model solution, as a limiting
   case.  We calculate the pre-main sequence (PMS) evolutionary tracks
   of a solar mass star in a dense stellar cluster environment and
compare
   it to
   one that forms in isolation. Up to an age of $950\,000\,$years
   differences in the accretion history lead to significantly
   different temperature and luminosity evolution. As accretion fades
   and the stars approach their final masses the two dynamic PMS tracks 
   converge. After that the contraction of the quasi-hydrostatic 
   stellar interiors dominate the overall stellar properties and
   proceed in very similar ways. Hence the position of a star in the
   Hertzsprung-Russell diagram becomes a function of age and mass
   only. However, our quantitative description of cloud fragmentation,
   star formation and early stellar evolution predicts substantial
   corrections to the classical, i.e.\ hydrostatic and initially fully
   convective models: At an age of 1 million years the proto-Sun is 
   twice as bright and 500 Kelvin hotter than according to 
   calculations that neglect the star formation process.
\end{abstract}

\keywords{Hertzsprung-Russel diagram --
hydrodynamics --
stars: evolution --
stars: formation --
stars: pre-main sequence -- 
Sun: evolution}



\section{Introduction}
Stars are born in interstellar clouds of molecular hydrogen.
Very often supersonic turbulence is observed in those clouds 
and then the mass growth of young stars are intimately coupled 
to the dynamical cloud environment. Stars form by gravitational 
collapse of shock-compressed density fluctuations(e.g.\ Elmegreen 
1993, Padoan 1995, Klessen, Heitsch, \& Mac~Low 2000, Heitsch, 
Mac~Low, \&  Klessen 2001, Padoan \& Nordlund 2001). But other
clouds are known, that are quiet and closely resemble 
isothermal, hydrostatic `Bonnor-Ebert'-spheres (Alves et al.\ 2001).

Once a gas clump becomes gravitationally unstable, it begins to
collapse giving birth to a protostar.  While the structure of
molecular clouds is well studied observationally, our knowledge about
intrinsic properties of these youngest stars relies almost entirely on
theoretical stellar models. These models give ages, masses and radii
when brightness, distance and effective temperature 
are known. The so
determined ages constitute the only practical `clock' for tracing the
history of star-formation regions and for studying the evolution of
circumstellar disks and planet formation.  They constitute the basis
of our empirical understanding of the evolution of the young Sun and
the origin of solar systems.
 
Until recently (Wuchterl \& Tscharnuter 2001, hereafter WT) it was
necessary to {\em assume} a set of initial conditions for the stars at
very young ages (typically at a few $10^5$ years) in order to
calculate the properties at larger ages.  Usually the internal thermal
structure of the star is estimated at a moment when the dynamical
infall motions from the cloud are thought to have faded and the
stellar contraction is sufficiently slow, so that pressure forces
balance gravity.  Then hydrostatic equilibrium is a good
approximation.
Young star properties are therefore usually calculated without
considering gravitational cloud collapse and protostellar accretion in
detail. See, however, Winkler \& Newman (1980a,b), but also Hartmann,
Cassen \& Kenyon (1997) for a discussion of the possible effects of 
accretion. Altogether, classical PMS calculations typically assume 
fully convective initial conditions as argued for by Hayashi (1961).

However, it can be shown that the assumption of fully convective
stellar structure {\em does not} result from the collapse of isolated,
marginally gravitationally unstable, isothermal, hydrostatic
equilibrium, so called 'Bonnor-Ebert' spheres (WT).  Hence, early
stellar evolution theory has to be reconsidered. Our novel approach
here is to determine the structure and evolution of young stars
following from a quantitative description of the cloud fragmentation
process (\S2).  We present the first calculation of the properties of
the new born star as being a member of a cluster of protostars forming
from the fragmentation of a highly-structured molecular cloud, and
follow in detail the collapse of a one solar mass fragment until it
becomes observable in the visible light (\S3). We demonstrate that the
newly born star shows the trace of the fragmentation and collapse
process during its main accretion phase and the early hydrostatic
PMS contraction. At an age of a million years, however, its properties 
are almost identical for quiet and turbulent cloud conditions (\S4).

\begin{figure*}[t]
\unitlength1cm
\begin{picture}(16, 5.4)
\put( 1.5, 0.0){\epsfxsize=15cm \epsfbox{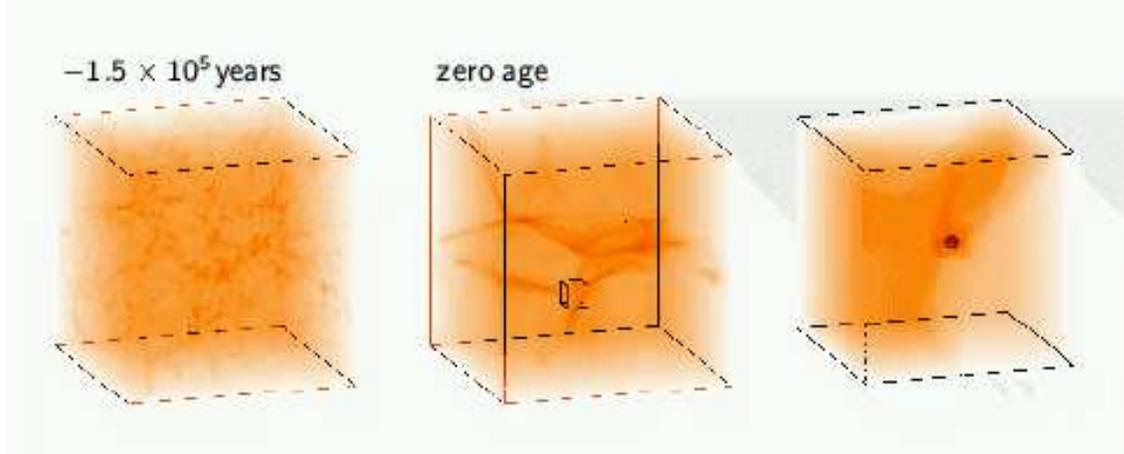}}
\end{picture}
\caption{The 3D density distribution
     of the dynamical molecular cloud fragmentation calculation at two
     different times (model $\cal I$ of KB00, KB01).  As we cannot
     treat the whole cloud, we focus on a sub-volume of mass
     $196\,$M$_{\odot}$ and size $(0.32\,{\rm{pc}})^3$. The left image
     depicts the initial random Gaussian fluctuation field, and the
     central image shows the system when the young proto-Sun reaches
     stellar zero age (i.e.\ when the cloud core for the first time
     becomes optically thick). During this $1.5\times10^5\,$year period
     local collapse occurs and a cluster of deeply embedded and
     heavily accreting pre- and protostellar condensations begins to
     build up (i.e.\ objects without and with central hydrostatic
     core).  The region where the proto-solar condensation forms is
     shown enlarged on the right-hand side. The volume considered in
     the 1D-RHD simulation is indicated by the circle.  }
\end{figure*}

\section{Towards a consistent description of star formation}
Molecular cloud complexes have diameters up to several 10$\,$pc and
masses in the range $10^3$ to $10^6\,$M$_{\odot}$.  They are
characterized by the presence of strongly supersonic turbulent gas
motions, as indicated by large observed line widths.  Without constant
energy input, interstellar turbulence dissipates rapidly on a
free-fall time-scale (Mac~Low et al.\ 1998, Stone, Ostriker, \& Gammie
1998, Mac~Low 1999).  Hence, the observed cloud structure is very
transient.  In the current calculation we consider a molecular cloud
region where turbulence is decayed and has left behind density
fluctuations characterized by a Gaussian random field which follows a
power spectrum $P(k) \propto 1/k^2$ (Klessen, Burkert, \& Bate 1998,
and in detail Klessen \& Burkert 2000, 2001, hereafter KB00 and KB01,
respectively).

The equation of hydrodynamics are solved using a particle based method
(SPH --- smoothed particle hydrodynamics, see Benz 1990, or Monaghan
1992) in combination with the special-purpose hardware device GRAPE
(GRavity Pipe, see Sugimoto et al.\ 1990, Ebisuzaki et al.\ 1993).  As
we cannot treat molecular clouds as a whole with sufficient numerical
resolution, we focus on a sub-region of the cloud with mass
$196\,$M$_{\odot}$ and size $(0.32\,{\rm{pc}})^3$ and adopt periodic
boundary conditions (Klessen 1997).  With a mean density of $n({\rm
H_2}) = 10^5\,$cm$^{-3}$ and temperature $T=10\,$K, the simulated
volume contains 222 thermal Jeans masses.  To be able to continue the
calculation beyond the formation of the first collapsing object,
compact cores are replaced by `sink' particles (Bate, Bonnell, \&
Price 1995) once they exceed a density of
$n({\rm{H}}_2)=10^9\,{\rm{cm}}^{-3}$, where we keep track of the mass
accretion, and the linear and angular momenta. The `sink' particle
size defines the volume of the 1D-RHD calculation.  The fragmentation
calculation covers a time-range from $-1.5\times 10^5\,{\rm years}$ to
$3.7\times10^7\,{\rm years}$.  The system is gravitationally unstable
and begins to form a cluster of 56 protostellar cores (as illustration
see Figure 1), corresponding to the `clustered' mode of star
formation (Klessen 2001b).

Besides that specific choice of the initial molecular cloud conditions, 
the only free parameters that remain in our dynamical star formation
model are 
introduced by the time-dependent convection-model, needed to describe 
stellar structure in a `realistic' way. These parameters are determined 
as usual in stellar structure theory by demanding agreement between the
model-solution and the actual solar convection zone as measured by
helioseismology.

We aim to describe the birth and the first million years of a solar
mass star. Therefore we select from the 3D cloud simulation that
protostellar core with final mass closest to $1\,$M$_{\odot}$ and use
its mass accretion history (Klessen 2001a, hereafter K01a) to
determine the mass flow into a spherical control volume centered on
the star. For the stellar mass range considered here feedback effects
are not strong enough to halt or delay accretion into this protostellar 
`feeding zone'.
%
%
Thus, the core accretion rates are good estimates for the actual stellar 
accretion rates. Deviations may
be expected only if the protostellar 
cores form a binary star, where the infalling mass must be distributed 
between two stars, or if very high-angular momentum material is
accreted, 
where a certain mass fraction may end up in a circumbinary disk and not 
accrete onto a star at all. For single stars matter accreting onto a 
protostellar disk may be temporarily `stored' in the disk before getting 
transported onto the star.
%
%
The flow {\em within} the control volume is calculated by solving the
equations of radiation hydrodynamics (RHD) in the grey Eddington
approximation and with spherical symmetry (Castor 1972, Mihalas \&
Mihalas 1984), in their integral form (Winkler \& Norman 1986).  At
any time we require that the total mass in the volume agrees with the
3D fragmentation calculation.  Convective energy transfer and mixing
is treated by using a new, time-dependent convection scheme (Wuchterl
\& Feuchtinger 1998) derived from the model of Kuhfu{\ss} (1987). We
include detailed equations of state and opacities (WT), and compute
deuterium burning processes with standard reaction rates (Caughlan \&
Fowler 1988) and convective mixing.  The 1D-RHD calculation covers a
spherical volume of radius $R = 160\, {\rm AU} =
2.46\times10^{15}\,{\rm cm}$ 
and contains a mass $0.028\, {\rm
M_\odot}$ at $t=0$. The calculation started $1.5\times10^5\,$ years
before the moment of stellar zero age, with a mass
$2\times10^{-5}\,{\rm M_\odot}$.  The final mass of the star is
$0.971\, {\rm M_\odot}$.

\begin{figure*}[t]
\unitlength1cm
\begin{picture}(16, 9.5)
\put( 2.0, 0.0){\epsfxsize=14cm \epsfbox{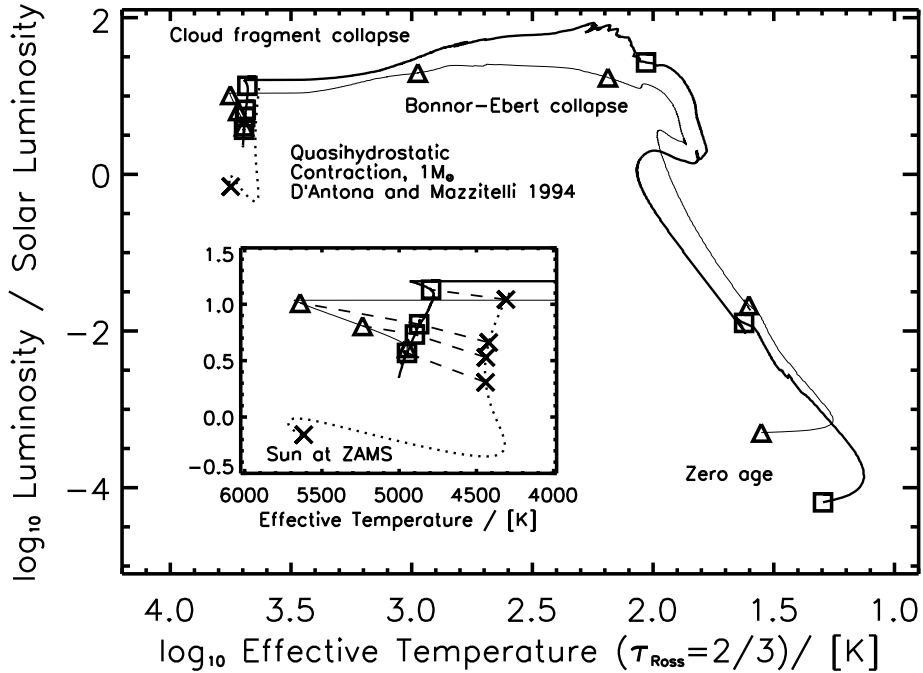}}
\end{picture}
\caption{Early stellar evolution in
             the Hertzsprung-Russell diagram.  Three evolutionary
             effective-temperature-luminosity relations (tracks)
             relevant to the young Sun are compared. The dotted line
             is a classical stellar structure, hydrostatic-equilibrium
             PMS track for $1\,{\rm M_\odot} $, for an initially fully
             convective gas sphere (`MLT Alexander' model of DM94). The
             two other lines are obtained by describing the formation
             of the star as a result of the collapse of an
             interstellar cloud.  The {\em thin line} is for a cloud
             fragment in initial equilibrium (a so called
             `Bonnor-Ebert' sphere of a solar mass, see WT for
             details). The (thick line) is for a cloud fragment that
             results from the dynamical fragmentation of a molecular
             cloud (KB00, KB01).  The two diamonds, in the lower
             right, indicate zero age for the two collapse-results.
             Triangles (WT, Bonnor-Ebert), squares (this work)
             and crosses (DM94) along the respective evolutionary 
             tracks mark ages of 1, 10, 100, $\rm 350\, kyr$, 0.5 
             and $\rm 1\,Myr$.  The cross at the end of the hydrostatic 
             track denotes the moment when energy generation by nuclear 
             reactions in the stellar interior, for the first time in 
             stellar live {\em fully} compensates the energy losses due 
             to radiation from the stellar photosphere, i.e.\ the zero 
             age main sequence (ZAMS). Corresponding age-marks for 0.1, 
             0.35, 0.5 and $\rm 1\,Myr$ are connected by dashed lines in
the 
             insert.
             }
\end{figure*}

\section{The formation of a $1\,$M$_{\odot}$-star}

The fragment we have chosen is highlighted in the 3D cloud structure
in Figure 1 at the moment when it becomes optically thick and departs
from isothermality, as determined by the 1D solution. We use this
instant to define the stellar zero age (WT).  The mass accretion rates
obtained for the selected fragment are strongly time varying and peak
around $1.5\times 10^{-5} \,$M$_{\odot}\,$yr$^{-1}$ (K01a).

Before reaching zero age, the temperature is close to the initial
cloud value of 10$\,$K and densities are still low enough so that the
heat produced by the collapse is easily radiated away from the
transparent cloud. Once the envelope becomes optically thick, the
temperature increases rapidly as the accretion luminosity rises. We
determine the effective temperature at the radius where the optical
depth $\tau_{\rm Ross} = 2/3$ (see Baschek Scholz, \& Wehrse 1991 for
a careful discussion).  Luminosity and temperature for the
non-isothermal phase obtained from the 1D-RHD calculation are shown in
Figure 2.  The zero age is marked by a diamond, close to the beginning
of the thick line in the lower right of Figure 2. For comparison the
results of a Bonnor-Ebert-collapse (WT) and a classical hydrostatic
stellar evolution calculation (D'Antona \& Mazzitelli 1994, hereafter
DM94) are shown as well. The equations used in the latter study
correspond to the hydrostatic limit of the current dynamical model
(WT), all physical parameters (opacities, etc.) are identical or
match closely (see WT).

The non-isothermal phase can be divided into three parts: (1) There is
a first luminosity increase up to $20\,{\rm L_\odot}$ with the
temperature staying below about 100$\,$K.  The central density of the
fragment rises until a hydrostatic core forms and the accretion flow
onto that core is accelerated 
until quasi-steady state is
established. (2) The subsequent main accretion phase leads to an
increase in temperature to 2000$\,$K while the luminosity shows a
broad maximum at $\sim 100\,$L$_{\odot}$.  Compared to the isolated
`Bonnor-Ebert case' the violent accretion in the cluster-environment
produces a considerably higher luminosity, and the oscillations around
maximum reflect the variable rates at which mass is supplied to the
accreting protostar as it travels through the dynamical environment of
the proto-cluster. (3) Once accretion fades, the star approaches
its final mass. The stellar photosphere becomes visible and the
luminosity decreases at roughly constant temperature. This is the
classical pre-main sequence (PMS) 
contraction phase, shown as a blow up in Figure
2.  The luminosity decreases at almost constant temperature and the
evolutionary tracks are nearly vertical being approximately parallel
to the classical `hydrostatic' track.

\section{Discussion}
Our dynamical model allows us to address the question of whether a
trace of the initial fragmentation and collapse process can be found
once the young star arrives at its final mass and becomes optically
visible.  Indeed, the star that forms in a dynamical cloud environment
is brighter when it reaches the PMS phase compared to the isolated
Bonnor-Ebert case. This is due to the higher accretion in the
dynamically evolving cluster environment (see also K01a).

As the mass accretion rates of evolving protostars in dense clusters
are influenced by mutual stochastic interactions and differ
significantly from isolated ones, the positions of stars in the main
accretion phase in the HR diagram are not functions of mass and age
alone, but also depend on the statistical properties of the
protostellar environment.  This affects attempts to infer age and mass
at this very early phase using bolometric temperatures and
luminosities of protostellar cores (see e.g.\ Myers \& Ladd 1993,
Myers et al. 1998). It is only possible 
as the statistical average over many different theoretical accretion
histories for different cluster environments or for an observational
sample of protostars with similar cloud conditions.

As the accretion flow fades away, however, the evolutionary tracks of
protostars {\em converge}, and the memory of environmental and initial
conditions is largely lost in the sense that one (final) mass 
corresponds to one track. For given mass and elemental composition
the stellar properties then depend on age only. For our one solar mass 
stars this happens at
$0.95\times10^6\,$years where the effective temperatures become equal
and remain within $20\,$K until the end of the 1D-RHD calculation at
$2\times10^6\,$years.  However, substantial {\em differences} remain
compared to the classical hydrostatic calculations.  The temperature
obtained from collapse models is consistently higher by about 500$\,$K
compared to classical hydrostatic computations at corresponding
luminosities.%
\footnote{ To indicate the consequences for stellar mass
determinations, we point out that during the second million years the
temperature of our one solar mass star corresponds to stars with
$2\,$M$_{\odot}$ on the classical hydrostatic tracks. The differences
in temperature and luminosity equivalently imply corrections for the
inferred ages: the `classical' luminosity at $10^6\,$years is
$6.2\,$L$_{\odot}$, while the corresponding `collapse' values are
smaller, $3.8\,$L$_{\odot}$ and $4.2\,$L$_{\odot}$, respectively. At
$2\times10^6\,$years, the new calculations give luminosities of twice
the solar value. The `classical' age for equivalent luminosities is
$0.8\times10^6\,$years.}

This deviation is the result of a {\em qualitatively} different
stellar structure (WT).  Most notable, the solar mass stars resulting
from collapse are {\em not} fully convective as is assumed in the
hydrostatic calculations, instead they do have a radiative core of
similar relative size as the present Sun. Convection is confined to a
shell in the outer third of the stellar radius. 
Stellar structure along the new dynamical
evolutionary tracks can be viewed as homologous to the present Sun
rather than to a fully convective structure.  Consequently, the
proto-Sun does not evolve along the classical Hayashi track for a
solar mass star, but roughly parallel to that.  It has a higher
effective temperature corresponding to the smaller radius of a
partially radiative object of the same luminosity as a fully
convective one.

As the dynamical PMS tracks converge for the two most extreme
assumptions about the stellar environment (dense stellar clusters vs.\
isolated stars) we predict that a solar mass star at
an age of $10^6\,$years will have a luminosity of
$4\pm0.4\,$L$_{\odot}$ and an effective temperature of
$4950\pm20\,$K. The uncertainties reflect the fading traces of the
adopted two highly disparate initial and environmental cloud
conditions.  For identical assumptions made about convection theory
and stellar opacities (DM94), the classical values are $2.0\,{\rm
L_\odot}$ and $4440\,$K, respectively.

Our predictions rely most critically on the correctness of the
following parts used in our argument: (1) The structure of young stars
can be calculated in spherical symmetry, (2) our prescription of
convection is sufficiently accurate outside the regime where it has
been tested, i.e.\ the present-day Sun, and (3) our radiative transfer
treatment and especially the present sources of stellar opacity are
sufficiently complete. However, it needs to be pointed out that {\em
all} those assumptions are also made for the classical calculations
that we have used for reference.

\acknowledgements We thank our second referee for comments and
discussion. This work was supported by the {\em
Nederlandse Organisatie voor Wetenschappelijk Onderzoek (NWO)}.
GW was partly supported by the {\em Fonds zur F\"orderung der
wissenschaftlichen
Forschung (FWF)} under project numbers S7305--AST and S7307--AST and
the {\em Deutsche Forschungsgemeinschaft (DFG)}, SFB 359 (key research
project of the German national science foundation on `Reactive flows,
diffusion and transport').   RSK
acknowledges support by a Otto-Hahn-Stipendium of the
Max-Planck-Gesell\-schaft and subsidies from a NASA astrophysics
theory program supporting the joint Center for Star Formation Studies
at NASA-Ames Research Center, UC Berkeley, and UC Santa
Cruz.
We thank the MPI f\"ur Astronomie, Heidelberg for use of their
GRAPE-System and E. Dorfi, Univ. Vienna for providing substantial
amounts of computational resources on his VAX-cluster where the 1D
computations were performed.


\end{document}